\documentclass[proof]{pasj01}
\draft
\usepackage{bm}
\usepackage{natbib}

\Received{}
\Accepted{}

 \usepackage{mathpazo}
 \usepackage[T1]{fontenc} 

\newcommand{\II}{II}
 
\begin{document} 

\title{ 
A numerical light curve model for interaction-powered supernovae}

\author{Yuki \textsc{Takei},\altaffilmark{1,2,3,}$^{*}$%
}
\altaffiltext{1}{Research Center for the Early Universe (RESCEU), Graduate School of Science, The University of Tokyo, 7-3-1 Hongo, Bunkyo-ku, Tokyo 113-0033, Japan}
\email{takei@resceu.s.u-tokyo.ac.jp}

\author{Toshikazu \textsc{Shigeyama}\altaffilmark{1,2}}
\altaffiltext{2}{Department of Astronomy, Graduate School of Science, The University of Tokyo, 7-3-1 Hongo, Bunkyo-ku, Tokyo 113-0033, Japan}
\altaffiltext{3}{Astrophysical Big Bang Laboratory, RIKEN, 2-1 Hirosawa, Wako, Saitama 351-0198, Japan}

\KeyWords{supernovae: general --- supernovae: individual (SN 2005kj, SN 2005ip) --- circumstellar matter --- stars: mass-loss --- shock waves}

\maketitle

\begin{abstract}
We construct a numerical light curve model for interaction-powered supernovae that arise from an interaction between the ejecta and the circumstellar matter (CSM).
In order to resolve the shocked region of an interaction-powered supernova, we solve the fluid equations and radiative transfer equation assuming the steady states in the rest frames of the reverse and forward shocks at each time step.
 Then we numerically solve the radiative transfer equation and the energy equation in the CSM with the thus obtained radiative flux from the forward shock as a radiation source.
We also compare results of our models with observational data of two supernovae 2005kj and 2005ip classified as type \II n and discuss the validity of our assumptions. 
We conclude that our model can predict physical parameters associated with supernova ejecta and the CSM from the observed features of the light curve as long as the CSM is sufficiently dense. Furthermore, we found that the absorption of radiation in the CSM is an important factor to calculate the luminosity. 

\end{abstract}

\section{Introduction}
Supernovae (SNe) show various properties in their spectra and light curves (LCs) strongly depending on how their progenitors have evolved.
Massive stars are known to lose their own envelopes due to the radiation pressure throughout their lives \citep*[e.g.,][]{CAK,2006ApJ...645L..45S}, leading to the formation of circumstellar matter (CSM) with a variety of density structures \citep[see][]{Smith2014}.
Especially, some progenitors seem to experience extremely intense mass-loss events shortly before the explosion, which results in the formation of dense CSM. If a massive star explodes as a SN in the CSM rich in hydrogen, some photons emitted from the SN are scattered by or ionize hydrogen atoms in the dense CSM and form narrow hydrogen emission lines in the spectrum.
Such a SN was classified as type IIn by \citet{Sch90} \citep[see also][]{1997ARA&A..35..309F}.
One of the highly important features of SNe IIn is an interaction between ejecta and CSM which dissipates the kinetic energy in the ejecta to the radiation energy.
In addition, the dense CSM may delay the emergence of a shock wave propagating in the envelope of the progenitor \citep[e.g.,][]{CI11} and in fact the delay of the shock breakout occurs in most SNe \II \ \citep{Forster2018}.
Therefore, we can extract some useful pieces of information about CSM from LCs of SNe IIn in the early brightening phases.
Studying LCs of SNe IIn is of significance in understanding the mass-loss history of progenitors, which may tell us about the final stage of the evolution of massive stars.
This will also help to test theoretical models for the evolution of massive stars shortly before the core-collapse.

Many investigations into LCs of SNe \II n have been done analytically, semi-analytically, and numerically  \citep[e.g.,][]{Ginzburg_Balberg,2012ApJ...746..121C,JPN2013,Dessart2015,DTF19}.
\citet{JPN2013} analytically derived the radius of the shocked region as a function of time and obtained bolometric LC of SN \II n  assuming that a cool dense shell is formed between SN ejecta and CSM due to the efficient radiative cooling.
\citet{Ginzburg_Balberg,DTF19} used the self-similar solution by \citet{Chevalier1982} for a shocked region, and conduct radiative transfer calculations in the shocked region and the unshocked CSM.
\citet{2012ApJ...746..121C} applied the model of \citet{A80} that presents analytical formulae expressing LCs of SNe \II-P and \II-L to SN \II n, and calculated the luminosity of SN \II n numerically.
\citet{Dessart2015} carried out multi-group radiation hydrodynamical calculations for SNe interacting with the CSM to investigate the origin of some types of super-luminous SNe. 

As these models do not resolve the shocked region efficiently or do not solve the structure by consistently taking into account the effects of radiation, here we present an alternative way to guarantee the spatial resolution in the shocked region. This method is not restricted to the power law density structures, which are necessary assumptions in some of the previous works mentioned above. 
We show that we can determine the distributions of physical quantities between the two shocks  as functions of time assuming a steady state in the rest frame of each of the shocks at each time step for given ejecta and the CSM structures.
Then we can calculate the LC by solving the radiative transfer equation and energy equation in the CSM ahead of the forward shock with the radiative flux from the forward shock as the inner boundary condition.

This paper is organized as follows;
In section 2, we formulate our model including the inner and outer boundary conditions.
Then we present results of our model, compare them with the observational data of SNe 2005kj and 2005ip, and discuss the validity of some of our assumptions in section 3.
Finally, we conclude the paper in section 4.

\section{Model}
After an explosion, a collision between SN ejecta and dense CSM  results in the formation of a forward shock propagating in the CSM and a reverse shock propagating in the SN ejecta.
This shocked region is composed of two components, shocked SN ejecta and shocked CSM, separated by a contact discontinuity.
In this section, we explain how we calculate spherically symmetric structures in these two shocked regions and derive the emergent radiative flux as a function of time and radius. Then we describe radiative transfer calculations in the unshocked CSM heated by the radiative flux emergent from the shocked region.
In the following, a subscript rs (fs) denotes physical quantities at the reverse (forward) shock.
\subsection{Shocked region}
Assuming a steady state in the rest frame of each of the shocks, we calculate the distributions of physical quantities in this region by integrating the following equations with respect to the radius $r$.
\begin{eqnarray}
&&\frac{\partial(r^{2}\rho v)}{r^{2}\partial r}=0, \\
&&v\frac{\partial v}{\partial r}+\frac{1}{\rho}\frac{\partial p}{\partial r}=0, \\
&&\frac{\partial}{r^{2}\partial r}\left[r^{2}\left\{\rho v\left(\frac{1}{2}v^{2}+e+\frac{p}{\rho}\right)+F\right\}\right]=0,
\end{eqnarray}
where $v$ denotes the velocity in the rest frame of each of the shocks, $\rho$ the density, $p$ the pressure, $e$ the specific internal energy, and $F$ denotes the radiative flux.
The velocity $v$ in the rest frame of the shock is transformed to $v+u$ in the rest frame of the center of the coordinate system ($r=0$), where $u$ is the shock velocity in the same frame.
Here we have assumed that the shocked region is in local thermodynamic equilibrium (LTE) state to avoid numerical instabilities associated with the integration. Thus we use the following equation of state,
\begin{equation}
p=\frac{\rho}{\mu m_{\rm u}}kT+\frac{1}{3}aT^{4},
\end{equation}
where $\mu$ is the mean molecular weight, $T$ the temperature, $m_{\rm u}$ the atomic mass unit, $k$ the Boltzmann constant, and $a$ the radiation constant.
$\mu$ becomes about 0.62 in fully ionized gases composed of the solar abundances of elements.
We fix $\mu=0.62$ in the shocked region.
Then the specific internal energy $e$ is given as
\begin{equation}
\rho e=\frac{3}{2}\frac{\rho}{\mu m_{\rm u}}kT+aT^{4}.
\end{equation}
Though this assumption of LTE is eventually broken at later epochs when the shocked region becomes optically thin and the radiation temperature could deviate from the gas temperature, we can assume LTE at earlier epochs  when the shocked region is optically thick. We will check the validity of this assumption for actual results presented in the next section.

When the shocked region is optically thick, we can apply the diffusion approximation to calculate $F$,
\begin{eqnarray}
F=-\frac{ac}{3(\kappa+\sigma)\rho}\frac{\partial T^{4}}{\partial r},
\end{eqnarray}
where $\kappa$ and $\sigma$ are the absorption coefficient and the scattering coefficient, and $c$ is the speed of light.
We use the Rosseland mean opacity presented by \citet{Ross1} that covers the temperature range of $3.75\leq\log T(\mathrm{K})\leq 8.70$ (OPAL opacity code). In the lower temperature range of $3.20\leq\log T(\mathrm{K})\leq3.75$, we use the opacity presented by \citet{Ross2}.

\subsection{Initial conditions and boundary conditions}
Assuming that homologously expanding SN ejecta have a density profile $\rho_{\rm ej}(r,t)$ with a broken power-law, one obtains
\begin{eqnarray}\label{eqn:density}
&&\rho_{\rm ej}(r,t)\nonumber\\
&&=\left\{
\begin{array}{l}
\frac{1}{4\pi(n-\delta)}\frac{[2(5-\delta)(n-5)E_{\rm ej}]^{(n-3)/2}}{[(3-\delta)(n-3)M_{\rm ej}]^{(n-5)/2}}t^{-3}\left(\frac{r}{t}\right)^{-n}\ (\frac{r}{t}\geq v_{t}), \\
\frac{1}{4\pi(n-\delta)}\frac{[2(5-\delta)(n-5)E_{\rm ej}]^{(\delta-3)/2}}{[(3-\delta)(n-3)M_{\rm ej}]^{(\delta-5)/2}}t^{-3}\left(\frac{r}{t}\right)^{-\delta}\ (\frac{r}{t}\leq v_{t}),
\end{array}
\right.
\end{eqnarray}
for a given ejecta mass $M_{\rm ej}$ and kinetic energy $E_{\rm ej}$ of the ejecta \citep{MM99}.
Here $v_{t}$ is determined so that the density be continuous at $r/t=v_{t}$.
The exponent $n$ depends on the envelope structure of the progenitor. The explosion of a blue super-giant (BSG)  (a red super-giant (RSG)) gives $n\simeq10$ ($\simeq12$).
The slope $\delta$ in the inner region takes a value in the range $0-1$.
We assume that the CSM density profile follows $\rho_{\rm CSM}\propto r^{-s}$.
We constrain $s<3$, because $s\geq3$ gives a solution in which the shock front is accelerating \citep{Chevalier1982}. When we use the density profile with $s=2$, which describes a steady mass-loss with a constant mass-loss rate and wind velocity, we set the wind velocity $v_{\rm w}$ to be $100\,\mathrm{km\,s^{-1}}$ for simplicity.
The Rankine-Hugoniot conditions relate physical quantities in the upstream (with subscript 0) of a shock wave to those in the downstream  (with subscript 1) by the following equations,
\begin{eqnarray}
&& \rho_{0}v_{0}=\rho_{1}v_{1},\\
&& \rho_{0}v_{0}^{2}+p_{0}=\rho_{1}v_{1}^{2}+p_{1}, \\
&& \frac{1}{2}v_{0}^{2}+e_{0}+\frac{p_{0}}{\rho_{0}}+\frac{F_{0}}{\rho_{0}v_{0}}=\frac{1}{2}v_{1}^{2}+e_{1}+\frac{p_{1}}{\rho_{1}}+\frac{F_{1}}{\rho_{1}v_{1}}.
\end{eqnarray}
As the mean free path of photons is much longer than that of gas particles, it is reasonable to assume that $F_{0}$ and $F_{1}$ take the same value.

In order to start the calculation of the inner structure of the shocked region from an initial time $t=t_\mathrm{ini}$, we first fix the initial radius of the reverse shock $r_{\mathrm{rs}}(t_{\rm ini})$ that can be obtained by the thin shell approximation \citep{JPN2013}.
The radius of the forward shock $r_{\mathrm{fs}}(t_{\mathrm{ini}})$ is determined so that the mass swept up by the forward shock matches the mass of the CSM  enclosed with the forward shock, i.e., the following equation 
\begin{eqnarray}
\int_{r_\mathrm{cd}}^{r_\mathrm{fs}}4\pi r^{2}\rho dr=\int_{0}^{r_\mathrm{fs}}4\pi r^{2}\rho_\mathrm{CSM}dr,
\end{eqnarray}
determines $r_{\mathrm{fs}}(t_{\rm ini})$. Here $r_\mathrm{cd}$ denotes the radius of the contact discontinuity satisfying
\begin{eqnarray}
\int_{r_\mathrm{rs}}^{r_\mathrm{cd}}4\pi r^{2}\rho dr=\int_{r_\mathrm{rs}}^{\infty}4\pi r^{2}\rho_\mathrm{ej}dr.
\end{eqnarray}
The velocities of the reverse shock and the forward shock, and the radiative flux at the forward shock front, are determined so as to satisfy boundary conditions at the contact discontinuity that the velocity, the pressure, and the radiative flux are continuous.
Then we can obtain a position $r_\mathrm{rs(fs)}(t+dt)$ at the next time step as $r_\mathrm{rs(fs)}(t+dt)=r_\mathrm{rs(fs)}(t)+u_\mathrm{rs(fs)}(t)dt$.

We stop the calculation when the temperature of a certain shocked region decreases down to 6000\,K, below which most of free electrons recombine.

\subsection{Radiative transfer calculations in the unshocked CSM}
Radiation emitted from the shocked region diffuses out in the dense unshocked CSM.
We solve radiative transfer equations in the unshocked CSM with the luminosities at the forward shock derived in the previous section as boundary conditions.
Although radiation may change the structure of matter through which it propagates, we ignore effects of such changes on the radiative transfer.
We carry out simplified two-temperature radiative transfer calculations described as below,
\begin{eqnarray}
&& \frac{\partial E}{\partial t}+\frac{\partial(r^{2}F)}{r^{2}\partial r}=4\pi\eta-\kappa\rho cE,\\
&& F=-\frac{c}{(\kappa+\sigma)\rho}\lambda\frac{\partial E}{\partial r},\label{eq:fld}\\
&& \lambda = \frac{2+R}{6+3R+R^{2}},\ R=\frac{|\partial E/\partial r|}{(\kappa+\sigma)\rho E},\\
&& \rho\left(\frac{\partial}{\partial t}+v_{\rm w}\frac{\partial}{\partial r}\right)U+\rho p_{\rm gas}v_{\rm w}\frac{\partial\rho^{-1}}{\partial r}=\kappa\rho cE-4\pi\eta,\\
&& 4\pi\eta=\kappa\rho acT_{\rm gas}^{4},
\end{eqnarray}
where we use the flux-limited diffusion approximation by \citet{LP81}. This formalism satisfies the condition that $|F|$ approaches to $cE$ in the optically thin limit and that equation (\ref{eq:fld}) converges to the diffusion approximation in the optically thick limit.  The specific internal energy of gas is denoted by $U=3kT_{\rm gas}/2\mu m_{\rm u}$, the gas pressure $p_{\rm gas}$, and $E$ is the radiation energy density, which can be expressed with a radiation temperature $T_\mathrm{rad}$ as $E=aT_\mathrm{rad}^4$.
Saha's equations shown below give us the mean molecular weight as $\mu^{-1}=X(1+n_{\rm HII}/n_{\rm H})+Y/4(1+n_{\rm HeII}/n_{\rm He}+2n_{\rm HeIII}/n_{\rm He})$ with the mass fractions  of hydrogen $X$ and helium  $Y$,
\begin{eqnarray}
\frac{n_{i+1}n_{\rm e}}{n_{i}}=\frac{2g_{i+1}}{g_{i}}\left(\frac{2\pi m_{\rm e}kT_{\rm gas}}{h^{2}}\right)^{3/2}\exp\left(\frac{-\chi_{i}}{kT_{\rm gas}}\right),
\end{eqnarray}
where $i$ denotes the ionization state associated with each element (here we consider gases composed of hydrogen and helium).
$\chi_{i}$ is the ionization energy and $g_{i}$ the statistical weight of ions in an ionization state $i$.
$n_{\rm e}$ and $n_{i}$ are the number densities of free electrons and ions of the ionization state $i$.
Initial conditions at time $t=t_\mathrm{ini}$ for the CSM are set as $T_{\rm gas}(r,t_{\rm ini})=2000\,\mathrm{K}$, $E(r,t_{\rm ini})=10^{-5}\,\mathrm{erg\,cm^{-3}}$, $F(r_{\rm fs}, t)=F_{\rm fs}(t)$, and $F(r>r_{\rm fs}, t_{\rm ini})=cE(r,t_{\rm ini})$.
It should be noted that these initial conditions of the unshocked CSM result in the initial luminosity of the order of $10^{37}\,\mathrm{erg\,s^{-1}}$, which is several orders of magnitude fainter than that from the shocked region.


\section{Results and discussion}
In this section we show results of our simulations with various parameter sets listed in table \ref{tab:my_label} and compare some of the resultant LCs with those from other models and well observed SNe \II n 2005kj and 2005ip. 
Throughout our work, we fix ejecta mass of $M_{\rm ej}=10M_{\odot}$ (where $M_{\odot}$ is the solar mass) otherwise mentioned.

\begin{table}[b]
  \tbl{Models and input parameters.}{%
  \begin{tabular}{cccccc}
      \hline
      \hline
      Model & $n$ & $s$ & $\delta$ & $E_{\rm ej,51}$\footnotemark[$*$] & $\dot{M}_{-3}$\footnotemark[$\dag$]  \\
      \hline
      E1M1 & 10 & 2 & 1 & 1 & 1\\
         E10M1 & 10 & 2 & 1 & 10 & 1\\
          E1M10 & 10 & 2 & 1 & 1 & 10\\
          E1M30 & 10 & 2 & 1 & 1 & 30\\
         E10M10 & 10 & 2 & 1 & 10 & 10\\ 
        E1M10n12 & 12 & 2 & 1 & 1 & 10\\
        SN 2005kj-a\footnotemark[\ddag] & 7 & 1.2 & 1.5 & 0.63 & 390 \\
        SN 2005kj-b & 7 & 1.2 & 1.5 & 1 & 390 \\
        SN 2005kj-c & 7 & 1.2 & 2 & 0.63 & 390 \\
        SN 2005kj-d & 7 & 2 & 1.5 & 0.63 & 390 \\
        SN 2005ip-a\footnotemark[\ddag]& 10 & 2.13 & 1 & 8.1 & 0.32 \\
      \hline
    \end{tabular}}\label{tab:my_label}
\begin{tabnote}
\footnotemark[$*$] Kinetic energy in units of $10^{51}\,\mathrm{erg}$.  \\ 
\footnotemark[$\dag$] Mass-loss rate in units of $10^{-3}M_{\odot}\,\mathrm{yr^{-1}}$. If $s\neq2$, the average values defined in equation (\ref{eqn:mass-loss}) are shown.\\
\footnotemark[$\ddag$]  Models for SNe 2005kj and 2005ip.
\end{tabnote}
\end{table}

\subsection{Shocked region}
\begin{figure*}[t]
    \begin{center}
    \includegraphics[width=0.9\linewidth]{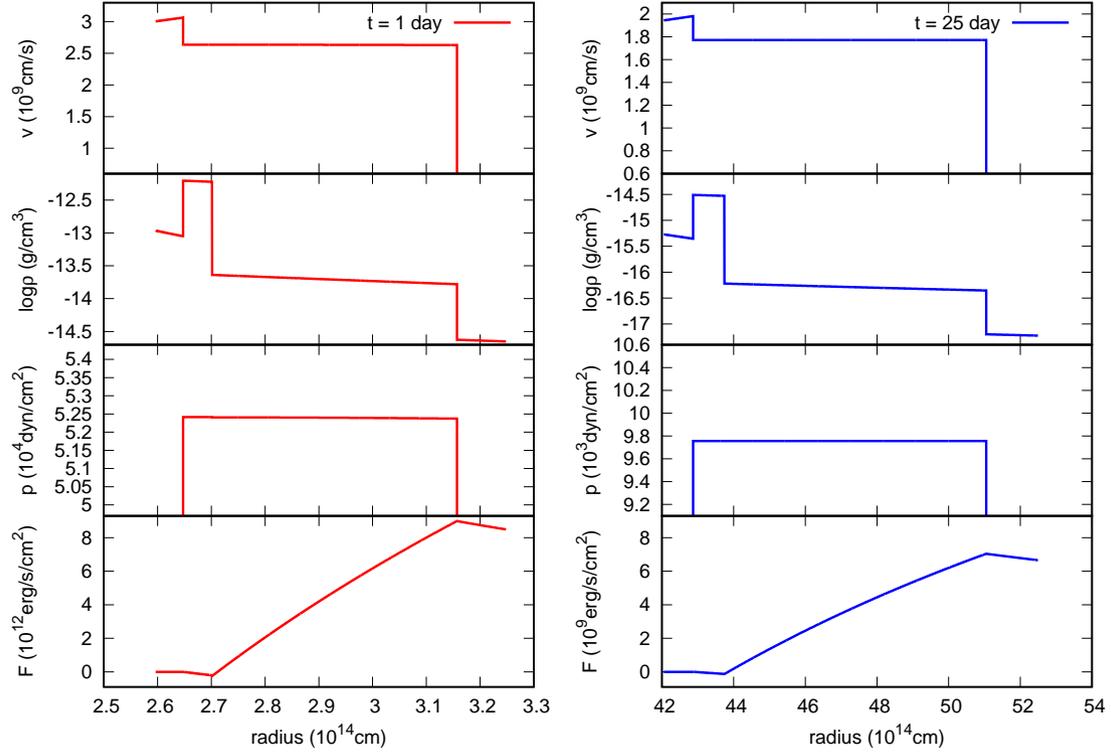}
    \end{center}
    \vspace{-5truemm}
    \caption{The structures around the shocked regions at 1 day (the left panels) and 25 day (the right panels) for model SN 2005ip-a. The profiles of velocity, density, pressure, and radiative flux are shown from the top panel to the bottom panel.
The reverse and forward shocks  at 1 day are located at $r\sim2.65$ and $\sim3.15\times10^{14}\,\mathrm{cm}$, respectively.
The radiative flux at $r>r_{\rm fs}$ follows $F\propto r^{-2}$ because the constant luminosity is assumed at $r>r_{\rm fs}$ in these particular plots.}
    \label{inner_t1}
\end{figure*}
\begin{figure}[t]
    \begin{center}
    \includegraphics[width=0.95\linewidth]{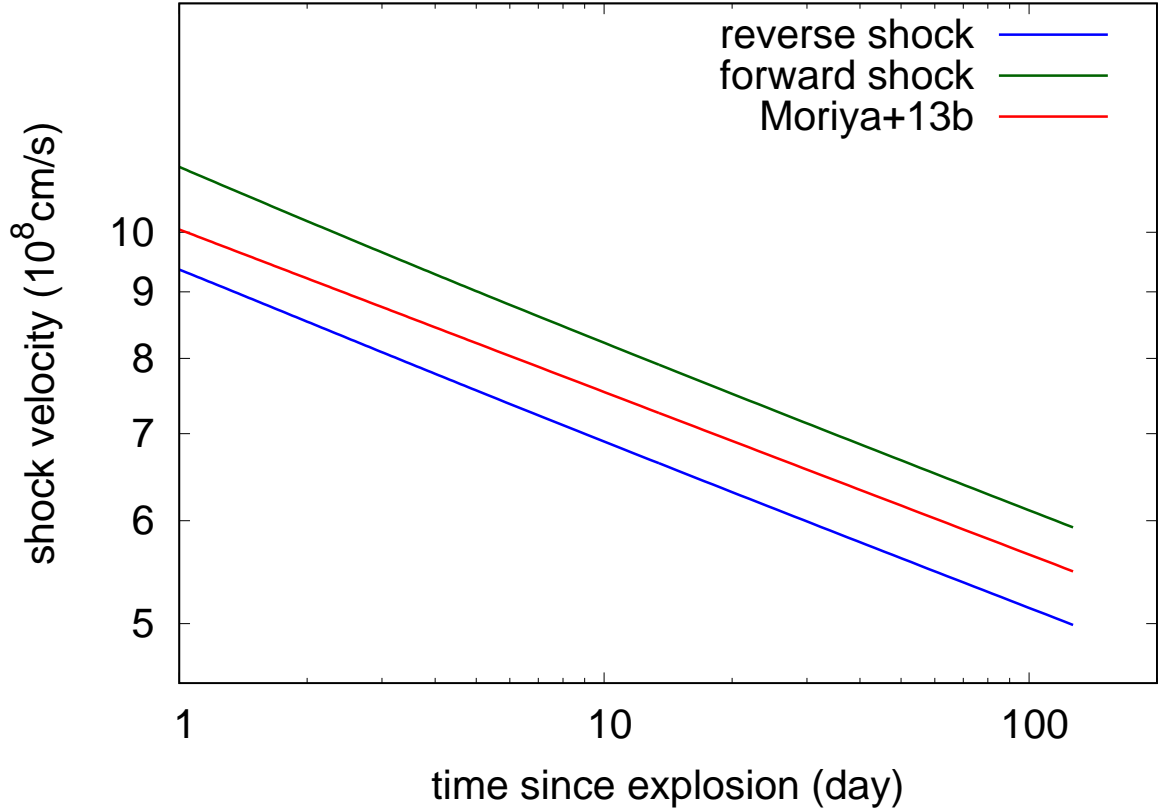}
    \end{center}
    \caption{Velocities of the reverse and forward shocks compared with the velocity of the thin shell derived in \citet{JPN2013} as functions of time for the same model in figure \ref{inner_t1}.}
    \label{fig:shock_velocity}
\end{figure}

Figure \ref{inner_t1} shows the inner structure of the shocked region for model SN2005ip-a in table 1 at 1 day and 25 day since explosion.
The width of the shocked region increases from $\sim5\times10^{13}\,\mathrm{cm}$ to $\sim8\times10^{14}\,\mathrm{cm}$ during this period while the contact surface moves from $2.7\times10^{14}\,\mathrm{cm}$ to $4.4\times10^{15}\,\mathrm{cm}$.
Thus the small fractional thickness of $\sim0.06$, which remains constant, suggests the validity of the thin shell approximation adopted in \cite{JPN2013}.
This is further supported from a comparison of the velocities of the reverse and forward shocks with the velocity of the thin shell derived by \citet{JPN2013} (Figure \ref{fig:shock_velocity}): The velocity of the thin shell runs between the velocities of the forward and reverse shocks, though the thin shell is less decelerated than the shocks.
This can be ascribed to radiative loss from the forward shock in our model.

\begin{figure}[t]
    \begin{center}
    \includegraphics[width=0.95\linewidth]{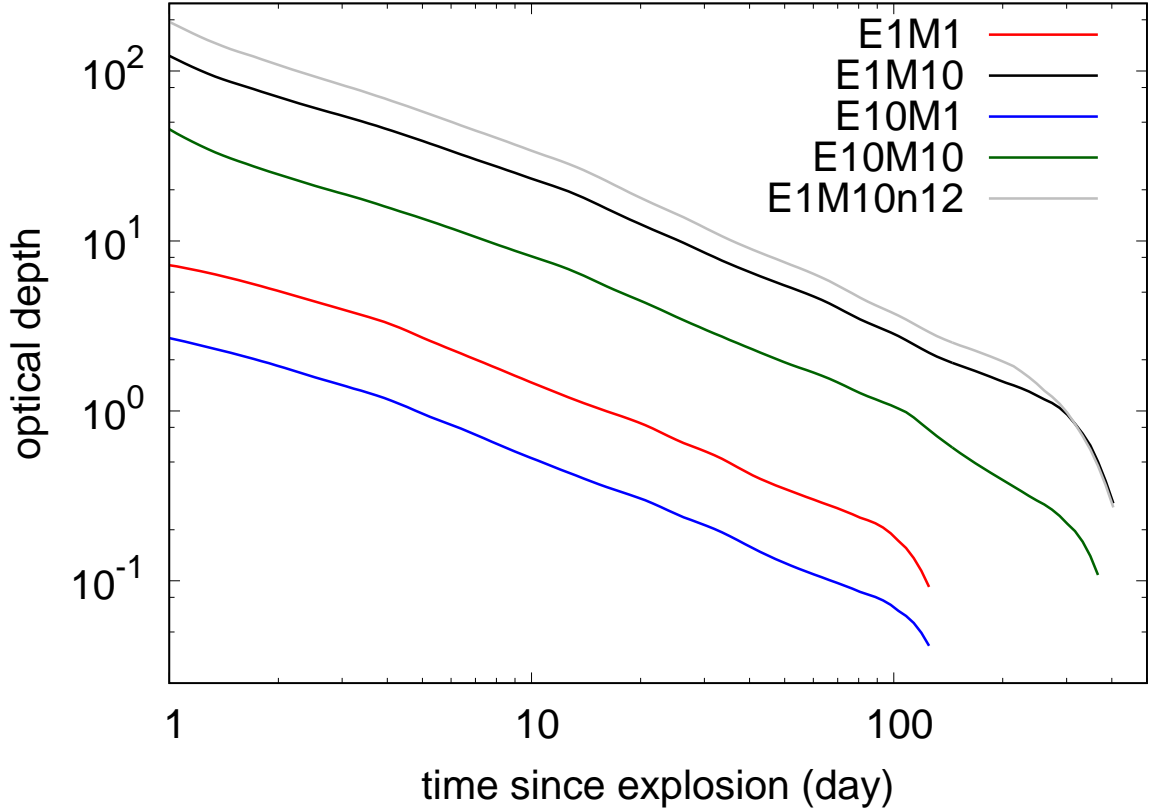}
    \end{center}
    \caption{Optical depth of the shocked region, measured from $r=r_{\rm fs}$ to $r=r_{\rm rs}$ for models indicated by the labels.}
    \label{fig:optical_depth}
\end{figure}

We also calculate the optical depth $\tau$ of the shocked region given by,
\begin{eqnarray}
\tau\equiv-\int_{r_{\rm fs}}^{r_{\rm rs}}(\kappa+\sigma)\rho dr,
\end{eqnarray}
and plot it as a function of time in figure \ref{fig:optical_depth}.
The shocked regions in models E10M10, E1M10, and E1M10n12 keep their optical depths greater than unity until $\sim100$ day, while those in models E1M1 and E10M1 becomes optically thin in a few days after explosion.
Since the scattering opacity is the dominant source of the total opacity, which is independent of density in high temperature regions, $\tau$ is roughly proportional to the density and width of the shocked region.
Figure \ref{fig:optical_depth} suggests that we can use the diffusion approximation until $\sim100$ days after explosion in the dense CSM but for only a few days in the less dense CSM.

\subsection{Conversion efficiency}\label{sec:ce}
\begin{figure}[t]
    \begin{center}
    \includegraphics[width=0.95\linewidth]{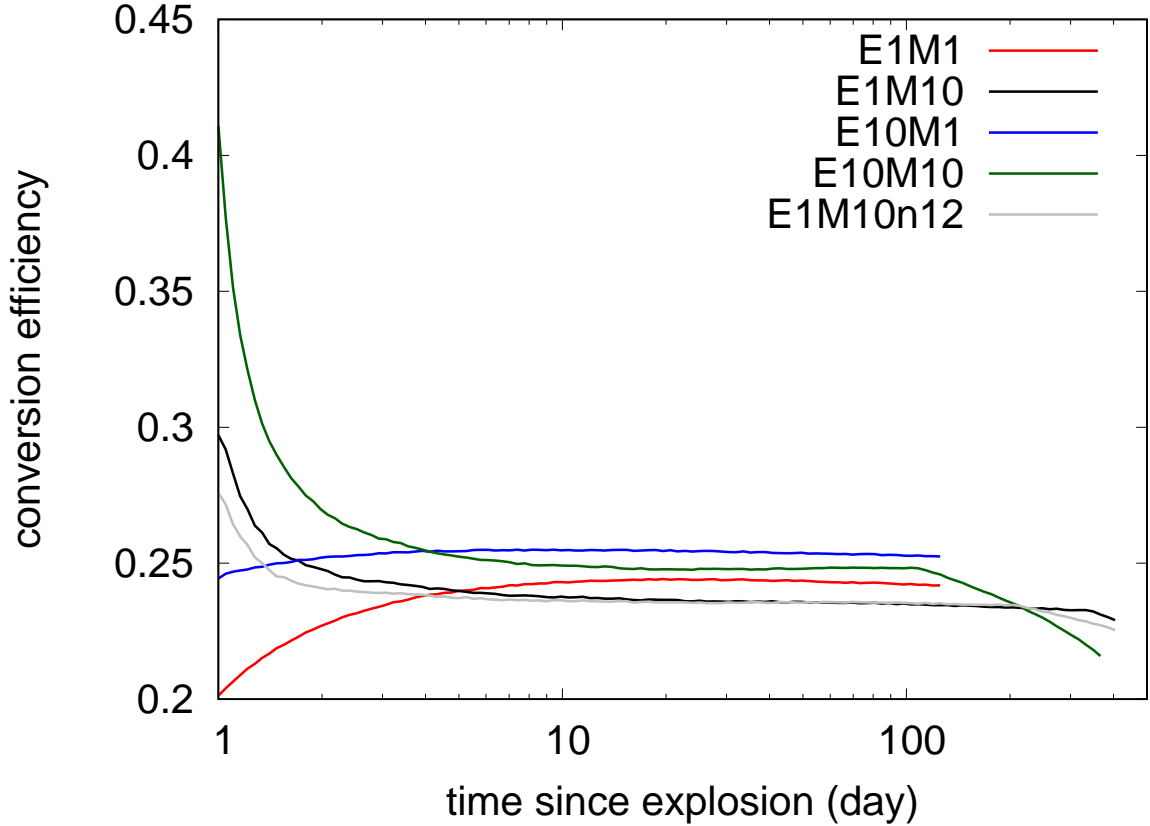}
    \end{center}
    \caption{Conversion efficiencies defined by Equation (\ref{eqn:conversion_efficiency}) for several models as functions of time.}
    \label{fig:conversion_efficiency}
\end{figure}

The shock waves dissipate a part of kinetic energy to radiation.
The conversion efficiency $\epsilon$ from kinetic energy to radiation can be defined as
\begin{eqnarray}
\epsilon=\frac{L}{dE/dt},
\end{eqnarray}
where $dE/dt$ is the kinetic energy incident to the shock front per unit time.
The efficiency $\epsilon$ has been roughly estimated to be $\sim0.1$ in the literature \citep{cf...2010MNRAS.407.2305V,JPN2013_1}.
The efficiency $\epsilon$ at the forward shock front is expressed as 
\begin{eqnarray}\label{eqn:conversion_efficiency}
\epsilon=\frac{F_{\rm fs}}{\frac{1}{2}\rho_{\rm CSM}u_{\rm fs}^{3}}.
\end{eqnarray}
The right hand side of the above equation can be obtained for our models and could be a function of time.
The conversion efficiencies of models listed in table \ref{tab:my_label} are plotted in figure \ref{fig:conversion_efficiency} as functions of time.
If a SN occurs in dense CSM as in models E10M10, E1M10, and E1M10n12, the conversion efficiency initially takes a high value and decreases for the first several days. Then the conversion efficiency almost keeps a constant value for the subsequent $\sim100$ days and abruptly drops at later epochs because the reverse shock enters into the inner ejecta where the density has a shallow profile $\rho_{\rm ej}\propto r^{-\delta}$ [see equation (\ref{eqn:density})]. The other models with less dense CSM exhibit different behaviors of the conversion efficiencies. The conversion efficiency gradually increases for the first several days. Then it keeps a constant value till the temperature in the shocked region drops below 6000 K. This is because the reverse shock still stays in the outer ejecta where the density has a steep profile.

\begin{figure}[t]
    \begin{center}
    \includegraphics[width=0.95\linewidth]{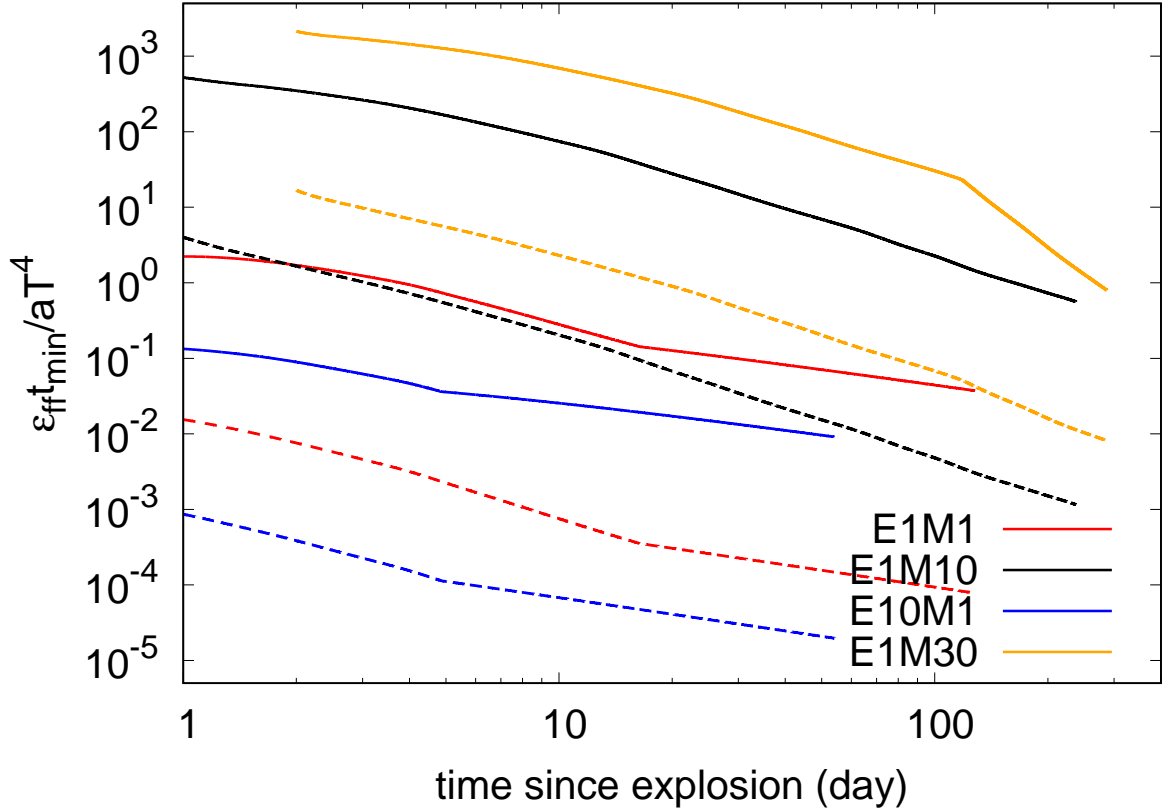}
    \end{center}
    \caption{The values of $\epsilon_{\rm ff}t_{\rm min}/aT^{4}$ at both shock fronts as functions of time for models E1M1, E1M10, E10M1 and E1M30.
    The solid lines show $\epsilon_{\rm ff}t_{\rm min}/aT^{4}$ at the reverse shock, while the dashed lines at the forward shock.
    Model E1M30 is calculated from day 2 while others from day 1.}
    \label{fig:LTE}
\end{figure}

Here we will discuss the validity of the LTE assumption by comparing the timescale to emit sufficient photons for thermal radiation estimated by $aT^4/\epsilon_{\rm ff}$ with  other characteristic timescales at the both shock fronts.  We consider the free-free emission whose emissivity $\epsilon_{\rm ff}\ [{\rm erg\,s^{-1}\,cm^{-3}}]$ depends on the density and temperature as 
\begin{eqnarray}
\epsilon_{\rm ff}(\rho, T_{\rm gas})\propto \rho^{2}T_{\rm gas}^{1/2},\label{eqn:eps_ff}
\end{eqnarray}
because photons are mainly generated by free-free emission at the both shock fronts.
If the gas in the shocked region attains the LTE state, the emitted radiation energy due to free-free emission exceeds the radiation energy density.
Thus $\epsilon_{\rm ff}(\rho, T_{\rm gas})t_{\rm min}/aT^{4}$ at the shock fronts can measure the validity of the LTE assumption, where $t_{\rm min}$ denotes the shortest characteristic timescale.
Since the shocked region changes due to the expansion and/or the diffusion of radiation, $t_{\rm min}$ can be estimated from 
\begin{eqnarray}
&& t_{\rm min}\equiv{\rm min}\{t,t_{\rm diff}\},\\
&& t_{\rm diff}\sim {\rm max}\{\tau, 1\}\frac{\Delta r}{c},
\end{eqnarray}
where $\Delta r\equiv r_{\rm fs}-r_{\rm rs}$ is the width of the shocked region.
When the total pressure is dominated by the gas pressure, we obtain the gas temperature $T_{\rm gas}$ as below, 
\begin{eqnarray}
&& T_{\rm gas,\ rs}\equiv\frac{3}{16}\frac{\mu m_{\rm u}}{k}(r_{\rm rs}/t-u_{\rm rs})^{2},\\
&& T_{\rm gas,\ fs}\equiv\frac{3}{16}\frac{\mu m_{\rm u}}{k}(v_{\rm w}-u_{\rm fs})^{2},
\end{eqnarray}
which are derived from the equation of state for monoatomic ideal gas.
Figure \ref{fig:LTE} shows $\epsilon_{\rm ff}(\rho, T_{\rm gas})t_{\rm min}/aT^{4}$ as functions of time for several models.
From this figure, the values of $\epsilon_{\rm ff}t_{\rm min}/aT^{4}$ at the reverse shock in models E1M10 and E1M30 exceed unity throughout their evolution while $\epsilon_{\rm ff}t_{\rm min}/aT^{4}<1$ in model E1M1 and E10M1.
Thus the LTE assumption holds at the reverse shock front in models E1M10 and E1M30.
On the other hand, the time dependence of $\epsilon_{\rm ff}t_{\rm min}/aT^{4}$ at the forward shock front shows different behaviors.
In models E1M10 and E1M30, $\epsilon_{\rm ff}t_{\rm min}/aT^{4}>1$ until a few days after the explosion, dropping below unity afterwards.
In model E10M1, we cannot assume the LTE state at both reverse and forward shock fronts throughout their evolution. If we impose the condition $\epsilon_{\rm ff}t_{\rm min}/aT^{4}>1$ holds at the forward shock for the first $100\times t_{100}$ days, a dimensional analysis implies that the criterion
\begin{equation}\label{eqn:crit}
\left(\frac{M_\mathrm{ej}}{M_\odot}\right)^{3/4}\left(\frac{E_{\rm ej}}{10^{51}\,{\rm erg}}\right)^{-3/4}\left(\frac{\dot{M}}{M_\odot\,{\rm yr}^{-1}}\right)\gtsim0.6t_{100},
\end{equation}
should be satisfied.
Thus our model can be applied to SNe \II n interacting with very dense CSM with a mass-loss rate $\gtsim 0.1\,M_\odot$ yr$^{-1}$ for a SN with $M_\mathrm{ej}\sim10\,M_\odot$ and $E_\mathrm{ej}\sim10^{51}$ erg, for example.

\subsection{Light curves}
\begin{figure}[t]
    \begin{center}
    \includegraphics[width=0.95\linewidth]{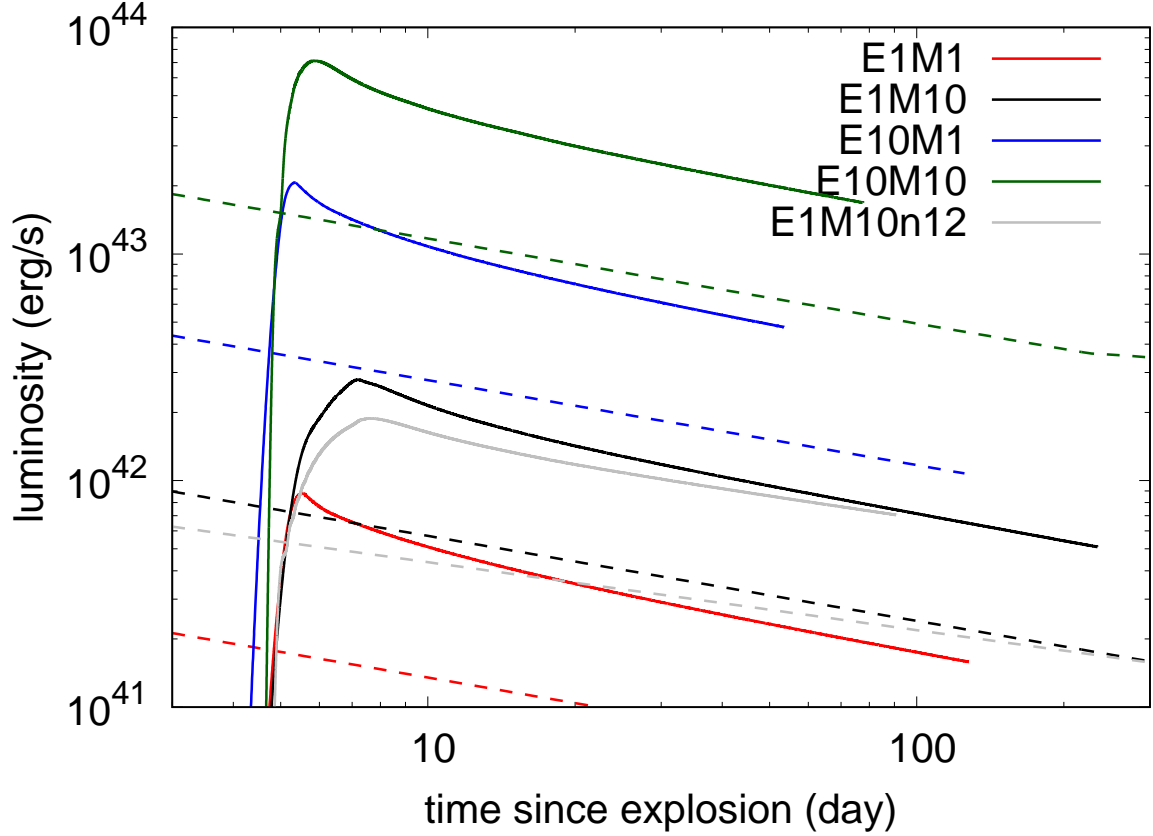}
    \end{center}
    \caption{Calculated LCs measured at the edge of the unshocked CSM (solid line).
    Dashed lines are the LCs by \citet{JPN2013}.
    Lines with the same color correspond to the same parameter set.}
    \label{fig:LC_cmp}
\end{figure}

Figure \ref{fig:LC_cmp} shows LCs calculated from the luminosity at the outer edge of CSM located at $r_{\rm out}=10^{16}\,\mathrm{cm}$.
 SNe \II n are brighten as photons emitted from the shocked region diffuse out from the CSM. Thus if we compare models E1M1 and E1M10, model E1M10 with denser CSM has a wider peak in its LC and a higher peak luminosity as shown in this figure.  More energetic ejecta result in an earlier time of the peak luminosity and a higher peak as seen from a comparison of models E1M10 and E10M10. More rapid expansion of the shocked region shortens the diffusion time of photons in the CSM. This is the reason why a factor of 10 energetic model has more than a factor of 10 higher peak luminosity. Afterwards, the luminosity declines following the same power law as that of \citet{JPN2013} until the reverse shock enters the inner ejecta.

\subsubsection{Comparison with \citet{DTF19}}\label{sec:DTF19}
\begin{figure}[t]
    \begin{center}
    \includegraphics[width=0.95\linewidth]{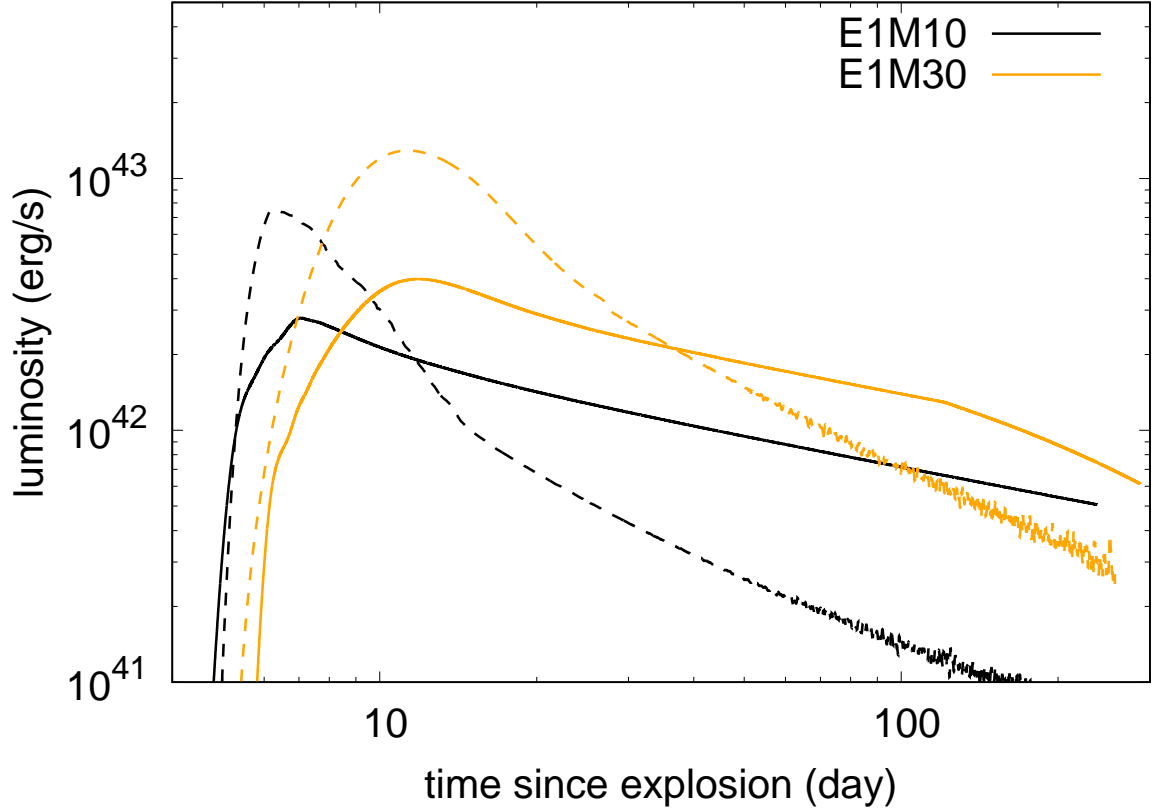}
    \end{center}
    \caption{Comparison of our models (solid line) with models by \citet{DTF19} (dashed line).
    The lines with the same color correspond to models with the same parameter set.}
    \label{fig:DTF19}
\end{figure}

To check the validity of assumptions made in our models, we compare our models E1M10 and E1M30 with other models in \citet{DTF19}, which used the same setup for the initial conditions.
In figure \ref{fig:DTF19}, we compare our LCs with those derived by \citet{DTF19}.
Models E1M10 and E1M30 have dimmer peak luminosities compared with \citet{DTF19}, while the initial rise time is similar. 
In our models, a part of radiation energy heats up the unshocked CSM and thus the luminosity is reduced. \citet{DTF19} did not take into account this effect. 
We will estimate how much radiation energy  $\Delta E$ is used to heat up the unshocked CSM  by the following formula.
\begin{eqnarray}
\Delta E\sim\int(L_{\rm fs}-L(r=r_{\rm out}))dt,
\end{eqnarray}
where $L_{\rm fs}$ denotes the luminosity emergent from the forward shock and calculated as $L_{\rm fs}=4\pi r_{\rm fs}^2F_{\rm fs}$ and $L(r)$ denotes the luminosity distribution.
We obtain values of $\Delta E\sim5.4\times10^{47},\,1.6\times10^{48}\,\mathrm{erg}$ for models E1M10 and E1M30, respectively.
If $\Delta E$ were emitted as radiation over a certain time $\Delta t$, peak luminosity $L_{\rm p}$ would increase by $\Delta E/\Delta t$.
If $\Delta t$ is characterized by the initial rise time, $\Delta E/\Delta t\sim2\times10^{42},\,3\times10^{42}\,{\rm erg\,s^{-1}}$ for models E1M10 and E1M30, respectively. These values could explain the difference between the models compared in figure \ref{fig:DTF19}.
Thus the heating of the CSM and resultant reduction of the emergent flux should be taken into account when calculating the luminosity around the peak. 

The assumption of LTE of our models overestimates the radiative flux after the region becomes optically thin as seen from the tails brighter than those of \citet{DTF19}.
From figure \ref{fig:LTE}, $\epsilon_{\rm ff}t_{\rm min}/aT^{4}$ at the reverse shock front exceeds unity in both models. 
By contrast, at the forward shock $\epsilon_{\rm ff}t_{\rm min}/aT^{4}\lesssim1$ in both models at later epochs as discussed in section \ref{sec:ce}, which also leads to the overestimation of the luminosity.
The luminosity of our model follows power-law evolution until the reverse shock enters the inner ejecta, while the luminosity of models by \citet{DTF19} shows steeper drop in the optically thin phase due to the more properly treated emission process.

\subsection{Comparison with observations}
Here we compare our results with two well observed SNe 2005kj and 2005ip. We show what kind of information we can obtain and discuss the limitation of our model.
\subsubsection{SN 2005kj}
We have picked up this SN because the mass-loss rate of the progenitor was estimated to be $\sim0.9\,M_{\odot}\ {\rm yr^{-1}}$ \citep{JPN2014}. This high mass loss rate indicates that this SN may satisfy the criterion (\ref{eqn:crit}).

\begin{figure}[t]
    \begin{center}
    \includegraphics[width=0.95\linewidth]{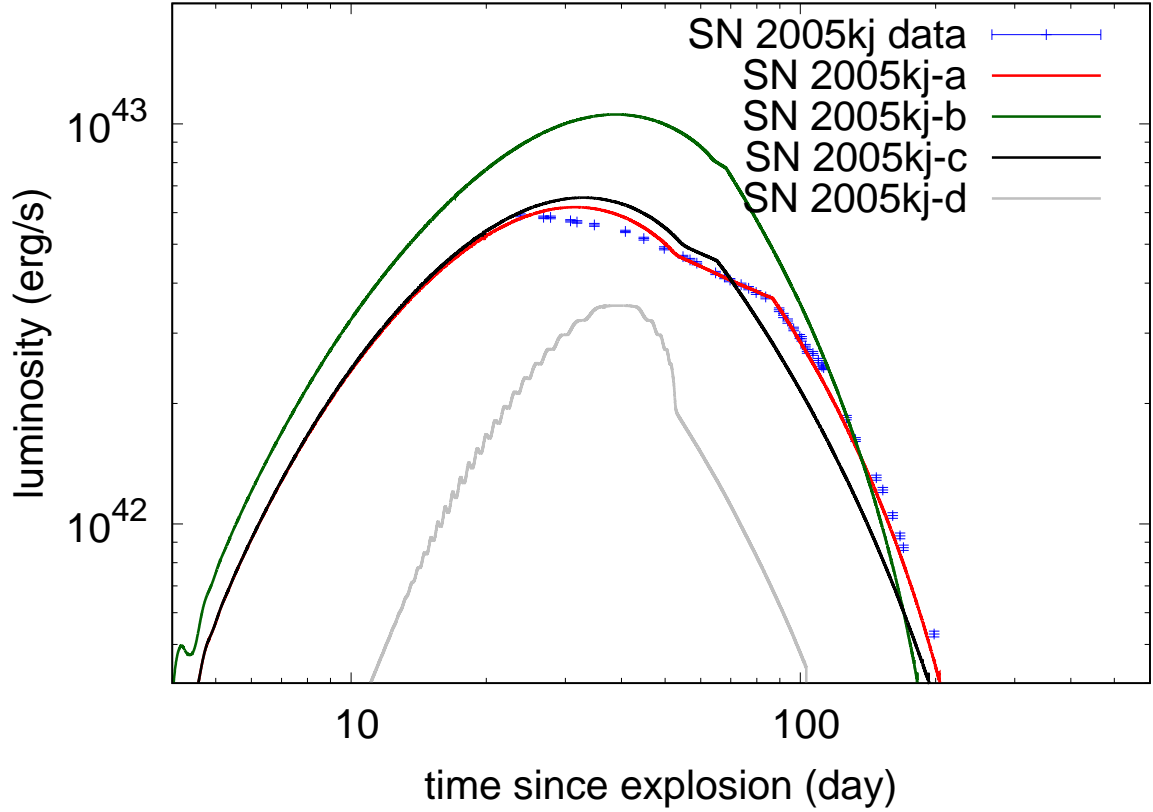}
    \end{center}
    \caption{Comparison of the observational data of SN 2005kj with our numerical models.
    }
    \label{fig:cmp_2005kj}
\end{figure}

To reproduce the bolometric LC of SN 2005kj constructed from optical and near-infrared observations \citep{T13_2013A&A...555A..10T} as shown in figure \ref{fig:cmp_2005kj}, we need to assume the explosion date 20 days before the discovery. The LC of this SN has a break at $t\sim100$ d (We will refer this epoch to $t_\mathrm{t}$). We found that most of the emission after this epoch comes from the shocks having already entered the inner core to reproduce the rapidly dropping flux. Thus the value of $\delta$ affects the shape of the LC and is found to be 1.5 to reproduce the observed LC. The best fit model (SN 2005kj-a) requires the other exponents to be $(n,\,s)=(7,\,1.2)$ and the energy of $E_{\rm ej}=6.3\times10^{50}\,\mathrm{erg}$.
The CSM density distribution of the best fit model is given by
\begin{eqnarray}
\rho_{\rm CSM}(r)\simeq9.7\times10^{-14}\left(\frac{r}{10^{15}\,\mathrm{cm}}\right)^{-1.2}\,\mathrm{g\,cm^{-3}}.\label{eqn:CSM_2005kj}
\end{eqnarray}
Since $s$ is not equal to 2, the required mass loss is not stationary.
We define the average value as
\begin{eqnarray}
\left<\dot{M}\right>&\equiv&\frac{v_{\rm w}}{r_{\rm out}}\int_{R_{\rm p}}^{r_{\rm out}}4\pi r^{2}\rho_{\rm CSM}dr \nonumber \\
&\simeq&\frac{4\pi Dv_{\rm w}}{3-s}r_{\rm out}^{2-s}\label{eqn:mass-loss},
\end{eqnarray}
where $R_{\rm p}\left(\ll r_{\rm out}\right)$ is the progenitor radius and $D$ denotes the proportional constant of the density of the CSM.
From equations (\ref{eqn:CSM_2005kj}) and (\ref{eqn:mass-loss}), we obtain the mean mass-loss rate of $\left<\dot{M}\right>\simeq0.39(r_{\rm out}/(5\times10^{15}\ {\rm cm}))^{0.8}\,M_{\odot}\, {\rm yr^{-1}}$.
This value is of the same order of magnitude as that derived by \citet{JPN2014}.
Such a high CSM density satisfies the criterion (\ref{eqn:crit}), thus the assumption of LTE state can be justified throughout the evolution of SN 2005kj shown in this figure.

Figure \ref{fig:cmp_2005kj} also shows the other three models SN 2005kj-b, c, d compared with the plot of the bolometric LC of SN 2005kj.
If we change $E_{\rm ej}$ from $6\times10^{50}\,\mathrm{erg}$ to $10^{51}\,\mathrm{erg}$ (model SN 2005kj-b) without changing the other parameters from model SN 2005kj-a, the break point $t_\mathrm{t}$ appears $\sim20$ days earlier than in model SN 2005kj-a and the luminosity drops faster at later times.
On the other hand, $\delta$ affects $t_\mathrm{t}$ and the luminosity at $t\geq t_\mathrm{t}$ while it does not change the luminosity at early phase so much.
Model SN 2005kj-c with a larger $\delta$ exhibits a shorter $t_\mathrm{t}$ and flatter luminosity at later times, as shown in figure \ref{fig:cmp_2005kj}.
Moreover, we change $s$ from 1.2 to 2 in model SN 2005kj-d while keeping the averaged mass-loss rate $0.39\,M_{\odot}\,\mathrm{yr^{-1}}$ and the other parameters the same, which yields the CSM density of
\begin{eqnarray}
\rho_{\rm CSM}(r)\simeq2.0\times10^{-13}\left(\frac{r}{10^{15}\ {\rm cm}}\right)^{-2}\,\mathrm{g\,cm^{-3}}.
\end{eqnarray}
As compared with model SN 2005kj-a, the luminosity of SN 2005kj-d is significantly fainter at all times as shown in figure \ref{fig:cmp_2005kj}. This is considered to be caused by absorption of radiation in the CSM, which can be seen from a difference in the column densities between these two models.
In order to confirm this, we compare the column density $N(s)$ of model SN 2005kj-d with that of model SN 2005kj-a, calculated as
\begin{eqnarray}
N(s)&\equiv&\int_{r_{\rm fs}(t_{\rm ini})}^{r_{\rm out}}\rho_{\rm CSM}dr \nonumber \\
&=&\frac{D}{s-1}\left[r_{\rm fs}(t_{\rm ini})^{1-s}-r_{\rm out}^{1-s}\right].
\end{eqnarray}
This equation yields $N(2)\simeq2\times10^{3}\,\mathrm{g\,cm^{-2}}$ while $N(1.2)\simeq3\times10^{2}\,\mathrm{g\,cm^{-2}}$, which is smaller than $N(2)$ by a factor of 10. This larger $N(2)$ indicates that more radiation from the shocked region is absorbed in the CSM rather than reaching the observer. 
For instance, the inner region of CSM, i.e. at $r=10^{14}\,\mathrm{cm}$ in model SN 2005kj-d, is an order of magnitude denser than in SN 2005kj-a while these are comparable at $r=10^{15}\,\mathrm{cm}$.
This means that much more radiation could be absorbed by this denser region and thus the emergent luminosity becomes fainter.


\subsubsection{SN 2005ip}
SN 2005ip is another well-studied SN \II n \citep[e.g.,][]{Smith_et_al_2009,2012ApJ...756..173S}.
The wind velocity and mass-loss rate are estimated to be $\sim100\,{\rm km\,s^{-1}}$ and $2.2\times10^{-4}\,M_{\odot}\,{\rm yr^{-1}}$ by \citet{Smith_et_al_2009}.
\citet{JPN2013} fitted a power law function:
\begin{equation}
    L\propto t^{\alpha}, \label{eqn:anl_lum_JPN}
\end{equation}
with an exponent $\alpha$ to the observed LC and  obtained $\alpha=-0.536$. This exponent is a function of $n$ and $s$ expressed as
\begin{equation}
    \alpha = \frac{6s-15+2n-ns}{n-s}.\label{eqn:JPN_alpha}
\end{equation}
Then they derived two parameter sets of $(E_{\rm ej,51},\, \dot{M}_{-3},\, n,\, s)=(13,\, 1.2,\, 10,\, 2.3)$ and $(15,\, 1.4,\, 12,\, 2.4)$ with $v_{\rm w}=100\ {\rm km\,s^{-1}}$ and $r_{\rm out}=10^{16}\,\mathrm{cm}$ for two different structures of the progenitor ($n=10$ and 12).
The temporal evolution of shock velocity and luminosity of our model at later epochs is almost the same as that of \citet{JPN2013}.

Thus, at first we fit the analytical model by \citet{JPN2013} to the observational data of SN 2005ip at later epochs and obtain $\alpha\simeq-0.447$. This is greater than the original value obtained in \citet{JPN2013} because we ignored the LC in the early phase ($t<30$ day) that is affected by the photon diffusion in the CSM.
From equation (\ref{eqn:JPN_alpha}), we obtain $s\simeq2.13$ when $n=10$.
Assuming the steady mass-loss ($s=2$), $n$ becomes about $n\simeq8.7$.
Furthermore, in order to restrict the parameters we fit the calculated shock velocity.
This procedure is the same as that of \citet{JPN2013}.
From evolution of the width of the H$\alpha$ profile, it is found that the shock velocity is $\sim1.8\times10^{4}\ {\rm km\,s^{-1}}$ at $\sim10-100\ {\rm day}$ \citep{2012ApJ...756..173S}.
Our best fit model suggests $E_{\rm ej}\simeq8.1\times10^{51}\,\mathrm{erg}$, which is smaller than $E_{\rm ej}\simeq1.3\times10^{52}\,\mathrm{erg}$ \citep{JPN2013} and
\begin{eqnarray}
\rho_{\rm CSM}(r)\simeq1.5\times10^{-18}\left(\frac{r}{10^{16}\,\mathrm{cm}}\right)^{-2.13}\,{\rm g\, cm^{-3}}.\label{eqn:CSM_2005ip}
\end{eqnarray}
Our model suggests a conversion efficiency higher than 0.1 and thus can reduce the kinetic energy of SN ejecta.
\begin{figure}[t]
    \begin{center}
    \includegraphics[width=0.95\linewidth]{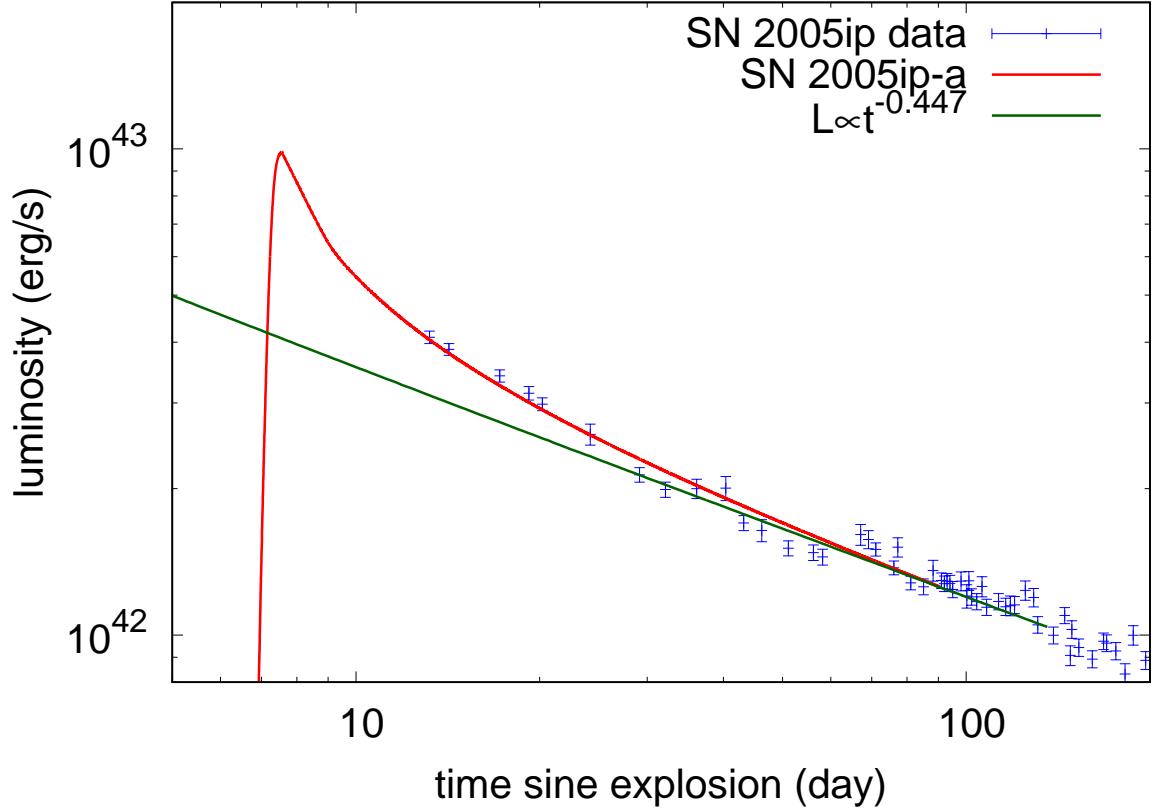}
    \end{center}
    \caption{Comparison of the observational data of SN 2005ip with our numerical model SN 2005ip-a.
    }
    \label{fig:CMP_SN2005ip}
\end{figure}
LCs with these parameters at $r=r_{\rm out},\ r_{\rm fs}$ are plotted in figure \ref{fig:CMP_SN2005ip} as well as the observational data of SN 2005ip.
From equations (\ref{eqn:mass-loss}) and (\ref{eqn:CSM_2005ip}) we obtain $\left<\dot{M}\right>\simeq3.2\times10^{-4}(r_{\rm out}/(10^{16}\,\mathrm{cm}))^{-0.13}\,M_{\odot}\,\mathrm{yr^{-1}}$. This low average value casts doubt on the validity of our assumption of the LTE according to the criterion given in formula (\ref{eqn:crit}). This means that we underestimated the mass-loss rate.
In fact, \citet{DTF19} estimated the mass-loss rate as $\left<\dot{M}\right>\simeq1\times10^{-2}M_{\odot}\,\mathrm{yr^{-1}}$ ($E_\mathrm{ej}=1.5\times10^{52}$ erg) from the LC fitting by their model, which partially takes into account the finite time to achieve the LTE. Thus we need to explicitly incorporate emission and absorption processes in the formulation for the shocked region as was done for the CSM to obtain physical parameters by comparison with this particular SN LC data with our model.

\section{Conclusion and future perspective}
We successfully constructed a model that guarantees to spatially resolve structures in the shocked region between SN ejecta and CSM, assuming a steady state in the rest frame of each of the shocks and calculate radiative transfer equations and energy equation to derive the luminosity at the outer edge of the CSM. We assumed the LTE in the shocked region to avoid numerical instabilities associated with integration with respect to the radial coordinate, while we explicitly included terms describing radiative emission and absorption in the CSM. By doing so, we can predict the peak luminosity of a SN \II n for thick CSM. Thus the structure of CSM can be inferred from observed initial rise times of the LC of a SN \II n.
As discussed in section \ref{sec:DTF19}, the unshocked CSM plays a crucial role in reducing radiative flux emergent from a forward shock front.

In the near future, we can test our model by a large number of observational data in the early phase of SNe \II n, or the peak luminosity, which will be detected by exceedingly wide-field and high-cadence optical camera Tomo-e Gozen and/or Zwicky Transient Facility \citep[e.g.,][]{Tomo-e-2016,ZTF_2014SPIE.9147E..79S}.
From these facilities we will be able to study the relationship between the initial rise times and the peak luminosities of SNe \II n.

We found that the assumption of LTE near the forward shock is broken in models with mass-loss rates often inferred from SNe \II n. In the next step we need to take into account radiative absorption and emission processes in the shocked region. This could be done if we succeed in suppressing numerical instabilities associated with the integration of the energy equation with respect to the radius in the shocked region. We will try an implicit method to integrate the energy equation to see if we can suppress the instabilities.

Though we assume spherical symmetry throughout the paper, many observations about the geometry of CSM have revealed that those of some progenitors of SNe \II n have aspherical structures  \citep[e.g.,][]{asphericity...2000ApJ...536..239L,asphericity...2008ApJ...688.1186H,2016ApJ...832..194K}.
\citet{2019arXiv191109261S} calculated 2D radiation hydrodynamic simulations for a spherical ejecta colliding with the circumstellar disk.
Our method may be applicable to this large scale asphericity by using aspherical CSM structures and/or aspherical ejecta structures. To do so, we need to include components of radiative flux and velocity in the other directions. This is rather straightforward, though the formulation becomes much more complicated. In addition, we need to treat the asphericity caused by turbulent motion of gas, which will develop to smaller scales. This requires a high spatial resolution and may weaken the feasibility of our method.



\begin{ack}
The authors thank Takashi J. Moriya for giving the observational bolometric light curve data of SN 2005kj and SN 2005ip to us.
The authors also appreciate that Daichi Tsuna gives us the results of his numerical simulations.
YT is supported by RIKEN Junior Research Associate Program.
YT is profoundly grateful to Daichi Tsuna for the discussion.
This work is partially supported by JSPS KAKENHI Grant Numbers 16H06341, 16K05287, 15H02082, MEXT, Japan.
\end{ack}

\bibliographystyle{apj}
\bibliography{IIn}

\end{document}